\documentclass[preprint,journal]{vgtc}            % preprint (journal style)

\onlineid{1080}

\preprinttext{To appear in IEEE Transactions on Visualization and Computer Graphics.}

\vgtccategory{Research}

\vgtcpapertype{please specify}

\title{LIVE-GS: LLM Powers Interactive VR Experience with Physics-Aware Gaussian Splatting}

\author{%
  \authororcid{Haotian Mao}{0009-0005-1294-4337},
  \authororcid{Hangyu Zhou}{0000-0003-3027-4523}, \authororcid{Zhuoxiong Xu}{0009-0006-7831-0051}, \authororcid{Siyue Wei}{0009-0001-9960-2719}, \authororcid{Yule Quan}{0009-0009-2938-0266}, \authororcid{Yan Zhang}{0000-0002-7549-4725},\\ \authororcid{Zixuan Guo}{0000-0002-0451-8988}, \authororcid{Nianchen Deng}{0000-0002-5292-266X} and 
  \authororcid{Xubo Yang}{0000-0001-5378-4003}
}

\authorfooter{
  \item
  	Haotian Mao, Zhuoxiong Xu, Siyue Wei, Yule Quan, Yan Zhang, Zixuan Guo and Xubo Yang are with Shanghai Jiao Tong University. E-mail: \{maohaotian | vector-02 | weisiyue071024 | qyl728 | yan-zh | zixuanguo | yangxubo\}@sjtu.edu.cn.
  \item
  	Hangyu Zhou is with Zhejiang University of Science and Technology. E-mail: zhy@zust.edu.cn.
  \item Nianchen Deng is with Shanghai AI Lab. E-mail: dengnianchen@pjlab.org.cn.
  \item Corresponding author: Xubo Yang.
}

\abstract{
 As 3D Gaussian Splatting (3DGS) emerges as a leading approach for novel view synthesis and scene reconstruction, its potential in digital asset creation has gained significant attention. An increasing number of asset libraries based on GS are being established. However, generating physics-based dynamic assets remains a time-consuming and expertise-intensive task, especially for non-experts. In this paper, we propose LIVE-GS, a highly realistic Virtual Reality (VR) system powered by Large Language Models (LLMs), which enables rapid creation of dynamic Gaussian assets and real-time VR interactions. To inform our system design, we conducted interviews to examine challenges faced by current GS-based VR systems and the specific demands of users. Based on these insights, we employed GPT-4o to analyze key physical properties of objects that significantly impact user interactions, ensuring physics-based interactions in VR align with real-world phenomena. A key innovation of LIVE-GS is its ability to predict reasonable parameters in just 10 seconds from static Gaussian assets while maintaining high-quality VR interactions. To validate our approach, we invited participants experienced in physical simulation to manually adjust physical parameters, providing a baseline for comparison in both asset quality and authoring efficiency. We also conducted a comprehensive user study to evaluate system usability and user satisfaction. Experimental results demonstrate that LIVE-GS, leveraging LLMs' scene understanding capabilities, can achieve efficient physical scene creation and natural interactions without requiring manual design or annotation.} % end of abstract


\teaser{
  \centering
  \includegraphics[width=1\linewidth]{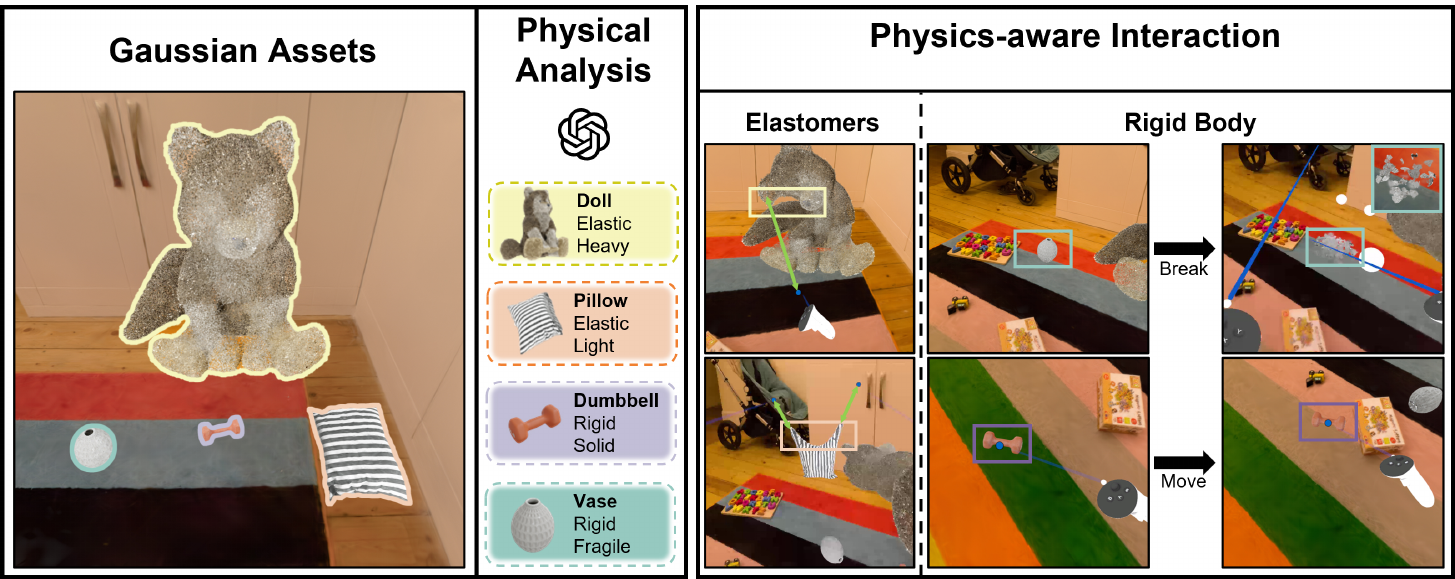}
  \caption{We introduce \textbf{LIVE-GS}, an innovative interactive VR system based on Gaussian Splatting powered by large language models (LLMs), which is capable of analyzing physical properties from static Gaussian assets within 10 seconds. Designed for human-centric VR immersion, our system supports real-time interactions and various physical simulations, including transformation, deformation, and fracture effects, thereby achieving an immersive VR experience for users.}
  \label{fig:teaser}
}

\graphicspath{{figs/}{figures/}{pictures/}{images/}{./}} % where to search for the images

\usepackage{tabu}                      % only used for the table example
\usepackage{booktabs}                  % only used for the table example
\usepackage{lipsum}                    % used to generate placeholder text
\usepackage{mwe}                       % used to generate placeholder figures

\usepackage{wrapfig}
\usepackage{subcaption}
\usepackage{graphicx}
\usepackage[export]{adjustbox}
\usepackage{multirow}
\usepackage{bm}

\usepackage{mathptmx}                  % use matching math font
\usepackage{amsmath}
\usepackage{cite}
\usepackage{enumitem}

\newcommand{\highlightRewrite}[1]{\textcolor{teal}{#1}}
\newcommand{\highlightA}[1]{\colorbox{yellow}{#1}}
\renewcommand{\highlightA}[1]{#1}
\newcommand{\highlightB}[1]{\colorbox{yellow}{#1}}
\newcommand{\highlightC}[1]{\colorbox{pink}{#1}}
\renewcommand{\highlightRewrite}[1]{#1} % comment this for hightlight
\begin{document}

%%%%%%%%%%%%%%%%%%%%%%%%%%%%%%%%%%%%%%%%%%%%%%%%%%%%%%%%%%%%%%%%
%%%%%%%%%%%%%%%%%%%%%% START OF THE PAPER %%%%%%%%%%%%%%%%%%%%%%
%%%%%%%%%%%%%%%%%%%%%%%%%%%%%%%%%%%%%%%%%%%%%%%%%%%%%%%%%%%%%%%%

\firstsection{Introduction}

\maketitle

Recent advancements in 3D Gaussian Splatting  (3DGS) \cite{kerbl20233d} have demonstrated remarkable potential in VR content authoring, primarily due to its superior rendering quality and streamlined production process. Unlike traditional modeling techniques, 3DGS requires only multi-view images of the target, making it accessible for non-experts to create high-quality VR assets. Besides, its seamless integration with modern rasterization pipelines enables real-time performance, meeting the stringent low-latency requirements of VR applications. Nowadays, 3DGS is poised to become a dominant representation standard, accompanied by a steady expansion of asset databases utilizing this format \cite{ma2024shapesplat}.

To enhance realism and immersion in virtual environments, it is necessary to achieve physics-based interaction. Fortunately, the explicit point cloud format of Gaussian kernels, enriched with geometry information, facilitates particle-based approaches for physical simulations\cite{xie2024physgaussian}. Notably, VR-GS \cite{jiang2024vr} has pioneered an interactive physics-based VR system with static Gaussian assets, showing its potential for immersive VR scene reconstruction. 

However, existing systems lack intelligent scene understanding and rely heavily on extensive user input to assign physical properties, resulting in inefficient asset creation. To investigate the limitations in current systems and identify user needs, we conducted interviews with domain experts. The results reveal that manual adjustments require relevant professional knowledge and significant time investment. Participants further emphasized the need for automated prediction of physical properties and expected instant feedback during authoring without sacrificing interaction quality. These findings highlight the necessity of a system that supports automated parameter prediction. Although recent techniques can recover detailed physical parameters with video-based techniques \cite{zhang2024physdreamer, lin2025omniphysgs,huang2025dreamphysics}, their assumptions, such as fixed density, limit their applicability in interactive scenarios. Besides, their slow convergence leads to delayed feedback that disrupts the creative workflow, and their high computational demands hinder scalability for large-scale authoring needs. While other works utilizing LLMs to analyze physical properties can provide relatively fast estimation, their system design and evaluation largely overlook the authoring and interaction requirements of users \cite{mao2025live,mao2025llm, zhao2024efficient, xu2024gaussianproperty}.

Based on these insights, we introduce LIVE-GS (\cref{fig:teaser}), a highly realistic VR system built on 3DGS that integrates physical property analysis. Our goal is to generate physics-aware objects rapidly with static Gaussian assets while ensuring satisfactory interactions in VR. The key idea behind LIVE-GS is leveraging the strong scene understanding capabilities of large language models (LLMs) such as GPTs for physical task analysis. Trained on massive datasets, these models possess the potential to infer detailed physical properties from visual inputs. In our work, we unlock this capability through a tailored module that leverages GPT-4o with prompt engineering and function tools, enabling structured reasoning for advanced physical analysis and driving our physics engine. This module is specifically designed to analyze images rendered from Gaussian assets, ensuring accurate extraction of physical parameters. To ensure real-time computational efficiency, we have adopted eXtended Position-Based Dynamics (XPBD) \cite{macklin2016xpbd} for elastic body simulation, while utilizing invisible mesh techniques for rigid body simulation. For the completeness and versatility of our system, we still integrate scene reconstruction, allowing users to generate customized assets through multi-view image inputs.

We primarily evaluated our system through a user study, in which we compared our system's automated outputs with manual adjustments performed by experienced users. The results demonstrate that our system can outperform major seasoned authors while maintaining high efficiency. To further validate the interaction quality and usability of our system, we built an immersive VR scene and invited users to engage with it in the second user study. Feedback on the reconstructed scenes is highly positive, with particular praise for the physical effects.

In conclusion, our major contributions include:
\begin{itemize}
% TODO: a specific system
\item An investigation to explore the necessity of automated parameter configuration and to uncover detailed user requirements to guide system design.
\item A GS-based VR system powered by LLMs, which integrates physical analysis capabilities. Our analysis module can predict the physical parameters of a static Gaussian asset within 10 seconds while maintaining interaction quality, substantially reducing the difficulty and time required for dynamic asset creation.
% \item User studies to validate the interaction quality and prediction efficiency, demonstrating that our system can effectively balance interaction quality with authoring speed while providing a satisfactory real-time VR experience.

\end{itemize}

\section{Related Work}

\subsection{3D Gaussian Splatting}
In the field of novel view synthesis, 3D Gaussian Splatting \cite{kerbl20233d} has recently emerged, achieving remarkable real-time rendering with splat-based rasterization. Subsequent work simultaneously expands its applicability and capabilities. Original 3DGS neglects the underlying geometry information, leading to redundant Gaussian kernels, which limit the rendering speed in large scenes.  \cite{lu2024scaffold, ren2024octree} are devoted to dynamically guiding the distribution and rendering of kernels, achieving high quality without sacrificing efficiency. Besides, it complicates the recovery of accurate surfaces (mesh) with irregular arrangement, where \cite{guedon2024sugar,wolf2024surface,geiger20242d,turkulainen2024dn,fan2024trim} import depth and normal supervision for surface-aligned results. For object-level segmentation, SAGD \cite{hu2024semantic} proposes a boundary-enhanced segmentation pipeline by lifting a 2D foundation model with training, while many other works supplement additional identity features \cite{dou2024cosseggaussians,silva2024contrastive,ye2023gaussian,cen2024segment3dgaussians,wang2024gaussianeditor} during the training procedure. However, void-like hole defects can be exposed after object removal, caused by a lack of information. A common way to tackle this problem is leveraging 2D inpaint guidance \cite{ye2023gaussian} for fine-training, and depth priors greatly benefit this procedure \cite{liu2024infusion,huang2023point}. Furthermore, this technique can be extended to dynamic scenes such as video reconstruction \cite{chu2024dreamscene4d,bae2024per}, deformation field reconstruction \cite{guo2024motion,duisterhof2023md,yang2024deformable,huang2024sc}, and diffusion-based content creation\cite{liang2024diffusion4d,yin20234dgen,ren2023dreamgaussian4d}. These jobs contribute to the accumulation of high-quality 3D assets \cite{ma2024shapesplat}.

\subsection{Scene Understanding}
Scene understanding is essential in 3D scene generation and reconstruction. Many researchers have investigated aligning visual and language presentation, such as CLIP \cite{radford2021learning}, DINO \cite{zhang2022dino}, and ALIGN \cite{jia2021scaling}. The development of these models has led to a breakthrough in scene understanding tasks. Many works \cite {kirillov2023segment,cheng2023tracking,ding2023pla,lu2023open,peng2023openscene,takmaz2023openmask3d} develop pre-trained promptable models handling various object detection and segmentation problems on arbitrary data. Recently, further research has thoroughly investigated multi-modal capabilities for various scene information such as video, text, and audio \cite{zeng2022socratic,yan2022videococa,girdhar2023imagebind,zhang2023video}.

Significant research efforts in the radiance field have been dedicated to advancing high-quality rendering and ensuring multi-view consistency. Feature field distillation has been well explored in some NeRF-related work such as \cite{zhi2021place} and \cite{siddiqui2023panoptic}. Another stream of methods obtains semantic embeddings from pre-trained models for open-vocabulary 3d scene understanding \cite{siddiqui2023panoptic,liao2024ov}. 3DGS can also obtain semantic embeddings via optimization in similar way \cite{shi2024language,zhou2024hugs,qin2024langsplat,liao2024clip,shi2024language}. Recent studies have explored LLMs' potential for automating physical scenarios \cite{mao2025llm,mao2025live}, achieving some preliminary progress. Inspired by previous research, we design an image-based material analysis strategy based on GPT-4o in our system, thus achieving physical simulations consistent with visual effects.

\begin{figure*}[!t]
    \centering
    \subfloat{
    \includegraphics[width=\linewidth]{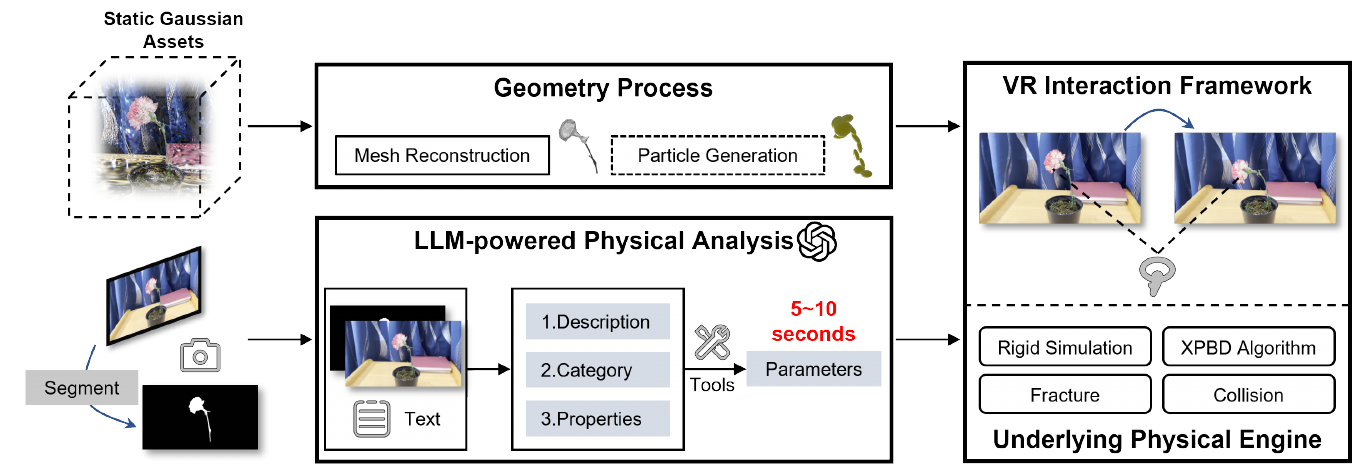}}
  \caption{\textbf{System overview}. Geometric information is extracted from static Gaussian assets via the \textit{Geometry Process}, generating discrete elements such as mesh or particles for physical simulations (\cref{subsection: simulation framework}). Two images, one rendered by the GS renderer and the other masked by SAM \cite{kirillov2023segment}, along with supplementary spatial information, are fed into the \textit{LLM-powered Physical Analysis} module, which directly outputs the required physical parameters (\cref{subsection: material analysis}). These results are integrated into our \textit{VR interaction framework} to deliver a fully immersive VR experience.}
  \label{fig: method overview}
\end{figure*}

\subsection{Physical Simulation in 3DGS}
% In the era of NeRF, its application in VR has been strictly hindered by its heavy rendering overload. Accordingly, quantities of work focuses on resolving real-time issues. Fov-NeRF\cite{deng2022fov} implements the first gaze-contingent neural radiance representation and foveated synthesis approach to extremely accelerate rendering speed. VR-NeRF \cite{xu2023vr} is tailored for high-fidelity VR content construction with multiple GPUs. 

In the era of NeRF, physics-based applications in VR have been strictly constrained by its implicit expression. Some works related to physics-based deformation \cite{qiao2022neuphysics,yuan2022nerf} rely on the extracted mesh. Although several researchers have achieved physical dynamics of complex particle systems \cite{guan2022neurofluid,li2023pac}, it is challenging to integrate them into VR applications due to the significant computational overload. Since the publication of 3DGS, its point cloud-based representation has opened new opportunities for various physics-based applications. PhysGaussian\cite{xie2024physgaussian} incorporates physical principles based on MPM into the rendering process, enabling realistic dynamics for different materials. For real-time simulation and interactions, VR-GS \cite{jiang2024vr} implements the first GS-based VR system based on XPBD, establishing a new paradigm for VR asset authoring. However, they still require users to possess professional expertise and invest significant time in manual adjustments. Although some learning-based methods \cite{zhang2024physdreamer,lin2025omniphysgs,huang2025dreamphysics} reconstruct the material field of static objects with the assistance of video generation models, their high computational overload and slow convergence hinder the responsive or large-scale authoring, and their underlying assumptions further limit applicability in real-time interactive scenarios. Recent works have also explored the possibility of using LLM for physical analysis \cite{mao2025llm,mao2025live,zhao2024efficient,xu2024gaussianproperty}, yet they overlooked users' practical demands when designing and evaluating their systems. In contrast, we propose LIVE-GS based on our interviews, which balances interaction quality with authoring efficiency.

\section{Preliminaries}

\subsection{Scene Reconstruction}

3DGS \cite{kerbl20233d} represents a scene using a collection of colored 3D Gaussians and renders them into camera views through splatting-based rasterization. Each 3D Gaussian primitive is characterized by its center $\mathbf{p}_k$ and covariance matrix $\Sigma_k$:

\begin{equation}
G_k=\exp{(-\frac{1}{2}(\bm{x}-\bm{p}_k)^T\bm{\Sigma}_k^{-1}(\bm{x}-\bm{p}_k))},
\end{equation}

where x is an arbitrary position in world space and the covariance matrix $\bm{\Sigma}=\bm{RSS}^T\bm{R}^T$ is formulated by scaling matrix $\bm{S}$ and rotation matrix $\bm{R}$. Additionally, each Gaussian kernel carries an opacity $\alpha$ and a color $c$ for tile-based rasterization. 3D Gaussian kernels $G_k$ are first transformed to image plane as 2D Gaussian kernels $G^{'}$\cite{zwicker2001ewa}, and then input for sorting and volumetric alpha blending:
\begin{equation}
    C=\sum_{i}^{N}c_i\delta_i\prod_{j=1}^{i-1}(1-\delta_j),    \delta_i = \alpha_iG_i^{'}(x^{'})
\end{equation}
where $x^{'}$ represents the target pixel, whose rendering result is influenced by N Gaussian kernels.

Starting from a set of images to produce a sparse point cloud with the SfM method\cite{snavely2006photo}, learnable parameters are optimized end-to-end through photometric loss. Furthermore, some work specialized in physical simulation, like \cite{xie2024physgaussian} introduces anisotropy loss in case over-skinny kernels generate prickly artifacts under deformations:

\begin{equation}
    \mathcal{L}_A=\frac{1}{|G|}\sum_{G_i \in G}max\{max(\bm{S}_{G_i})/min(\bm{S}_{G_i}),r\}-r
\end{equation}

which constrains the ratio between the maximum and minimum axes in the scaling parameter. Accordingly, total image loss becomes:
\begin{equation}
    \mathcal{L}_I = (1-\lambda_1)\mathcal{L}_1+\lambda_1 \mathcal{L}_{SSIM}+\mathcal{L}_A
\end{equation}

\subsection{Object Segmentation}
 An excellent segmentation can help build Gaussian asset datasets from the real world. Previous work has tried to separate interactive objects from the real world. For example, Gaussian Grouping \cite{ye2023gaussian} deploys SAM \cite{kirillov2023segment} to generate masks automatically, takes DEVA \cite{cheng2023tracking} to obtain consistent mask labels across training views as ground truth, and then introduces grouping loss, including \textbf{2D identity loss} and \textbf{3D regularization loss} for enhanced robustness and accuracy:
\begin{equation}
    \mathcal{L}_G=\lambda_{2d}\mathcal{L}_{2d} + \lambda_{3d}\mathcal{L}_{3d}
\end{equation}

With predefined threshold $\sigma_1$, each kernel $k$ of object $i$ is segmented through a classifier $C$:
\begin{equation}
    \{k|Softmax(C(k,i))>\sigma_1\}
\end{equation}

Other methods, such as \cite{hu2024semantic}, import a voting strategy as a post-process, which projects each kernel to screen space and determines whether its center is inside the corresponding mask. These help in extracting assets from entire reconstructed scenes.

\section{Interview: Investigating the Practical Requirements of Authors}
VR asset authoring for physics-based interaction can differ from traditional physical simulation tasks, yet this distinction has received limited attention in prior work. To bridge this gap and ground our system in real needs, we conducted targeted interviews before initiating its design. We interviewed 7 participants, 3 females and 4 males, aged 21 to 29 (\textit{M}=24.14, \textit{SD}= 2.64). All participants were experts in physical simulation tasks and had prior experience with 3DGS techniques and VR applications. Participants were first asked: \textit{1. What are the current challenges when you create interactive VR assets with 3DGS? 2. What factors contribute to these challenges?} According to their answers, we further asked three focused questions: \textit{3. Does the difficulty of parameter adjustment vary across different simulation phenomena? If so, how? 4. What are your expectations regarding simulation accuracy? 5. How do you prioritize time efficiency when creating assets?} 

\subsection{Finding 1: Current Challenges of asset authoring} 

All participants highlighted that manual adjustment of physical parameters is a cumbersome task. According to their practical experience, this process requires specialized expertise in materials science and physical simulation, which can impede users without such backgrounds. Furthermore, fine-tuning parameters to achieve desirable results tends to be time-intensive, as creators often need to perform repeated iterations to identify optimal parameter combinations. Worse still, as the number of interactive objects in a scene increases, the time required for parameter adjustment typically scales non-linearly, which arises from inter-object interactions and coupling effects. Authors often re-refine previously completed parameter setups to ensure consistency across the entire scene.

These insights indicate that manual parameter adjustment remains both dependent on specialized expertise and labor-intensive. This observation underscores the necessity of automation in physical asset creation.

\subsection{Finding 2: Authors' Demands for the Automated System} 

In this interview, participants reported that, regarding Q3, the difficulty indeed varies across simulation phenomena, largely due to differences in the number of parameters and the complexity of their implementation. For instance, configuring parameters for the fluid and the elastomers typically requires more effort than for a rigid body. For Q4, participants noted that their accuracy requirements depend on task objectives: High precision is necessary for industrial computing, whereas approximate but convincing results are usually sufficient for VR interactions. As for Q5, they emphasized that time efficiency typically involves a trade-off with accuracy, while highlighting a shared view that instant feedback can enhance the authoring experience under satisfactory interactive conditions.

\subsection{Summary} The previous interview reveals the value of automated parameter prediction. Tailored to VR asset authoring, our system should account for variations across different simulation types, ensuring that complex phenomena can be effectively analyzed. Without compromising VR \textbf{interaction quality}, it should aim to achieve \textbf{higher prediction efficiency} to foster an engaging authoring experience.

\section{LLM-powered VR System with Gaussian Splatting}
Building on the insights from our interviews, we present LIVE-GS, a GS-based VR system that supports automated parameter configuration and real-time interaction (\cref{fig: method overview}). We focus on two phenomena in our simulation framework to showcase the system's capabilities: elastic and rigid dynamics. To balance interaction quality with prediction efficiency, we developed an LLM-powered analysis module, further strengthened by prompt engineering and function-tool integration for more reliable reasoning.

\subsection{Real-time Physical Simulation Framework} \label{subsection: simulation framework}
\subsubsection{Mesh Generation}
We adopt mesh as an intermediate representation for subsequent physical simulations due to its algorithmic adaptability. We implement the alpha shape algorithm \cite{edelsbrunner1994three} to reconstruct the mesh from Gaussian kernels. In elastic body simulation, we reconstruct high-resolution meshes to better guide the generation of particles. For rigid body simulation, however, since meshes directly participate in the basic simulation and fracture, we choose a coarser reconstruction to improve efficiency. In this case, voxel downsampling is applied to the kernels, and a larger alpha value is used. As these meshes are not rendered at runtime, the reduced resolution remains sufficient.
\begin{figure}[htbp]
    \centering
    \includegraphics[width=1.0\linewidth]{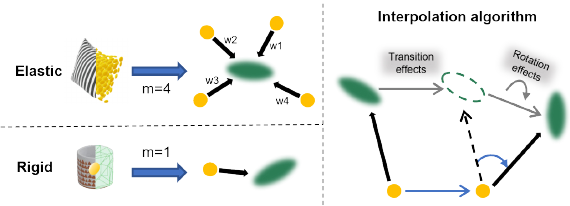}
    \caption{\textbf{Particle interpolation for Gaussian kernels.} The rotation of the particles not only contributes to the rotation quaternion but also affects the transition vector.}
    \label{fig:particle interpolation}
\end{figure}
\subsubsection{Elastic Body Simulation} 
While the Material Point Method (MPM) is often preferred due to its precision \cite{xie2024physgaussian}, its computational load hinders real-time VR applications. Consequently, we adopt XPBD for simulating elastic deformations. 

To mitigate the computational burden associated with the massive Gaussian kernels, we implement an interpolation scheme with invisible physical particles. These particles are generated from the mesh at initialization and participate in physical computations, strictly adhering to the physical constraints. Inside the object, we apply voxel sampling with density parameter $d$, filling one particle at the center of each voxel to provide structural support and prevent inward collapse. On the boundary, particles are placed at mesh vertices to preserve anisotropic geometric features, with a distance constraint requiring that any two particles remain farther apart than half the diagonal length of a voxel.

The states of these physical particles can be interpolated to determine the transformation of each Gaussian kernel, as illustrated in \cref{fig:particle interpolation}. For physical particles, changes in rotation and transition relative to initial states can be recorded as $\Delta \bm{q}$ and $\Delta \bm{t}$. For each Gaussian kernel, we identify the closest $m$ particles and calculate their weights $\bm{w}$ at frame 0, where $m$ is phenomenon-specific. In our work, the weights are computed based on the inverse of the initial distances between the kernel and particles, with normalization preserved. Accordingly, we track each particle’s rotational and translational deviations from its initial pose. The translation vector $\bm{t}'$ is obtained through linear interpolation, while the rotation is computed via recursive spherical linear interpolation:
\begin{equation}
    % T=\sum_i^{m} w_i (X_{i,r}^{p}-X_i^{p})
    \bm{t}^{'}= \sum_{i=1}^{m-1} w_i\Delta \bm{t}_i \quad \bm{q^{'}} = \mathrm{SLERP}_{\bm{w}}(\Delta \bm{q}_i).
    \label{eq:PS1}
\end{equation}
 
% \end{equation}
Additionally, the rotation of particles contributes to the translation vector, computed as:
\begin{equation}
    \bm{t_{i}}=\bm{x_{0}^{k}}-\bm{x_{i,0}^{p}},
    \label{eq:PS2}
\end{equation}
where $\bm{t_{i}}$ represents the initial distance vector between the kernel $\bm{x}_0^k$ and its i-th associated particle $\bm{x}_{i,0}^{p}$, precomputed during initialization to enhance rendering efficiency. The translational motion $\bm{t_q}$ induced by rotation can be calculated as:
\begin{equation}
    \bm{t}_r = \sum_{i=1}^{m} w_i(\Delta \bm{q}_i \bm{t}_{i} (\Delta \bm{q}_i)^{-1} - \bm{t}_{i}).
    \label{eq:PS3} 
\end{equation}
Finally, the runtime position $\bm{x}^k$ and rotation $\bm{r}^k$ can then be calculated as:
\begin{equation}
    \bm{x}^k = \bm{x}_0^k+\bm{t}^{'}+\bm{t}_{r} \quad \bm{r}^k = \bm{r}^{'}\bm{r}_0^k.
    \label{eq:PS4}
\end{equation}
We chose $m=4$ in our implementation to balance the efficiency and quality.  

\subsubsection{Rigid body Simulation} Rigid body simulation is relatively straightforward, as each Gaussian kernel maintains a fixed pose in its local coordinate system. Distinct from the particle-based elastic simulation, our rigid simulation algorithm operates directly on mesh representations, and each kernel's global pose can be derived by the transformations of the local coordinate applying \cref{eq:PS1,eq:PS2,eq:PS3,eq:PS4} with $m = 1$. To enhance the interactivity of rigid bodies, we further implement fracture behavior for fragile objects based on the Voronoi algorithm \cite{aurenhammer1991voronoi}, where the mesh is subdivided and splits into separate pieces when the external force exceeds a threshold. Each kernel is rebound to the closest new mesh's coordinate system when broken. Similarly, the pose of Gaussian kernels can be calculated using \cref{eq:PS1,eq:PS2,eq:PS3,eq:PS4}.

\subsection{LLM-powered Modules for Physical Analysis} \label{subsection: material analysis}

From the perspective of user experience, our physical analysis module is designed to analyze the parameters that have the most significant impact on interaction. We select GPT-4o as our base model due to its strong visual perception capabilities. We carefully prompt it to comprehend our requirements, reason as an expert author, and output a formatted reply, typically in a \textit{json} file, which seamlessly interfaces with our physics engine.
\highlightRewrite{\subsubsection{Input}
Our input consists of input images accompanied by supplementary textual information. Given the critical role of environmental context in our analysis, we provide two images: the original scene and a corresponding mask highlighting the target for analysis. Additionally, we supplement spatial information such as the camera's pose, the target's position, and the distance between them to enhance the model's spatial understanding. When analyzing multiple objects, the context from previous objects is preserved as a reference to ensure consistency across all targets.
\subsubsection{Parameters and Speculating}
Considering the complexity of our analysis task, we decompose it into sequential substeps, guiding our module to perform a step-by-step reasoning \cite{kojima2022large}. Upon receiving the images, the module first describes the content to extract the semantic information, subsequently inferring its constituent materials. Afterwards, it determines the target's category as rigid or elastic and further analyzes relevant parameters. For complex parameters, the module can better reason toward accurate results by first identifying the range and then analyzing specific values. To activate the capabilities of LLMs in physical analysis, we enable web search functionality to obtain richer knowledge for better understanding. When analyzing multiple assets in the same scene, the prior analysis is packed as contextual references.} 

Our simulation framework accepts parameters within specific ranges, which can be categorized as follows:
% \begin{enumerate}[(1)]
%     \item Elastomer, which exhibits deformation under stress and returns to its original shape when the stress is removed.
%        \begin{compactitem}
%          \item \textit{Mass}. 
%          \item \textit{Deformation Resistance}. 
%          \item \textit{Plasticity}. 
%        \end{compactitem}
%    \item Rigid body, characterized by negligible deformation under applied forces.
%        \begin{compactitem}
%          \item \textit{Mass}.
%          \item \textit{Fragility}.
%        \end{compactitem}
% \end{enumerate}

\begin{enumerate}[label=(\arabic*)] % enumitem 的标准语法
    \item Elastomer, which exhibits deformation under stress and returns to its original shape when the stress is removed.
       % itemsep=0pt 实现了 compactitem 的紧凑效果
       \begin{itemize}[noitemsep] 
         \item \textit{Mass} ($(0, \infty)$, g)
         \item \textit{Deformation Resistance} ($[0,1]$)
         \item \textit{Plasticity} ($[0,1]$)
       \end{itemize}
    \item Rigid body, characterized by negligible deformation under applied forces.
       \begin{itemize}[noitemsep]
         \item \textit{Mass} ($(0, \infty)$, g)
         \item \textit{Fragility} ($\{0, 1\}$)
       \end{itemize}
\end{enumerate}

\textbf{Mass}. As a fundamental physical attribute shared by both simulation modalities, mass plays a critical role in force dynamics and collision behaviours. Prior approaches, such as Physdreamer \cite{zhang2024physdreamer}, assume a uniform density across all Gaussian assets, thus leading to perceptually inaccurate weight representations for scene objects. To address this limitation, our module infers the target's size according to its environment context and then estimates the mass in grams.

\textbf{Deformation Resistance}. \highlightRewrite{In our XPBD algorithm, we add a shape matching constraint \cite{muller2005meshless}, which adjusts the predicted pose under forces of each particle towards the goal pose at each substep. The deformation resistance $d_R$ determines the extent to which predicted position $\bm{x}^p_{pre}$ is aligned with the goal position $\bm{x}^p_{goal}$:
\begin{equation}
    \bm{x}^p_f=(1-d_R)\bm{x}^p_{pre}+d_R \bm{x}^p_{goal}.
\end{equation}
}\highlightRewrite{This parameter indicates the stiffness of the elastomer, which is conceptually analogous to Young's modulus. Therefore, our module analyzes Young's modulus $E$ (MPa) of the target as a reference. However, the actual stiffness is also affected by the structural characteristics. For example, a laminated structure (such as pillows) or a helical structure (such as springs) can exhibit reduced resistance. To account for this, our module also estimates a structural coefficient $s$ to refine estimation:
\begin{equation}
    E^*=s E.
\end{equation}
This value is ultimately mapped to the range $[0,1]$ with following piecewise function:
\begin{equation}
    d_R=f(E^*)=a_i + b_i(E^*-E_i^b),\quad E_i^b\le E^* \leq E_{i+1}^b.
\end{equation}
This piecewise approach matches the reasoning logic of first determining the range. The intervals are set with several representative  materials, for which we carefully assign the \textit{deformation resistance} and predict related values with LLM. The intervals are set as $\bm{a}=[0,0.1,0.5,0.8,1]$ and $\bm{E^b}={[0,0.1,1,50,200]}$, where $i\in \{0,1,2,3\}$. $E^*$ is clamped between 0 and 200 before input.
}

\textbf{Plasticity}. Plastic deformation can occur under extreme stress, transitioning the object from elastic to plastic behavior, \highlightRewrite{which is controlled by input parameter $p$. In our implementation, we maintain a plastic deformation matrix $\bm{D}_p$ record the plastic state. At each simulation step, the stretch component $\bm{S}$ of the particle cluster is decomposed from the elastic deformation matrix, and the Frobenius norm $||\bm{S}-\bm{I}||_F$ can quantify the deformation magnitude, where $\bm{I}$ denotes the identity matrix. $\bm{D}_p$ is updated when $||\bm{S}-\bm{I}||_F$ exceeds the threshold $(1-p)$:
\begin{equation}
    \bm{D}_p^{new}=(\bm{I}+p(\bm{S}-\bm{I}))\bm{D}_p^{old},
\end{equation}
which represents the absorption of elastic deformation into the plastic component. $||\bm{D}_p-\bm{I}||_F$ is clamped to the range $[0,p]$ to prevent excessive deformation.} 

\highlightRewrite{We provide some classic samples with manual numerical values for few-shot learning. The LLM is required to compare the target against the references to determine the plasticity level, then infer the specific value within that identified range.
}

\textbf{Fragility}. As a binary parameter, fragility determines fracture simulation applicability, with fracture thresholds governed by simulation settings. When activated, the object undergoes fracture simulation based on the predefined threshold. 

Obviously, except for the universal parameter \textit{mass}, elastic object analysis presents greater complexity compared to the straightforward binary classification of \highlightRewrite{\textit{fragility}}, which explains why our subsequent experiments emphasize elastic dynamics more. The prompt template and more details can be found in the Appendix.
\begin{figure*}[!ht]
    \centering
    \subfloat{
    \includegraphics[width=\linewidth]{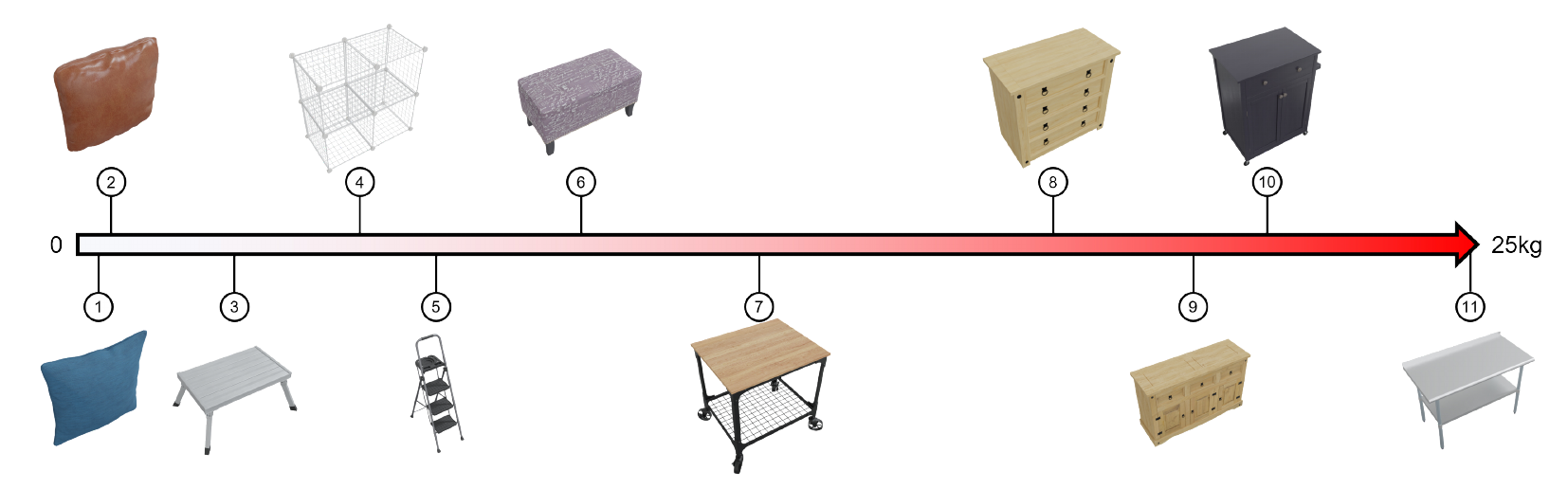}}
  \caption{\textbf{The mass estimation dataset}. We chose 11 objects ranging from 0.45kg to 25kg, ranking their mass in ascending order above.}
  \label{fig: mass truth}
\end{figure*}
\subsubsection{Function Tools} GPT-4o frequently commits computational errors due to its limitations in mathematics. \highlightRewrite{To address this issue, we leverage its capability to invoke external functions for computational assistance. We implement the essential computational functions mentioned above, package them as callable tools, and integrate them with the model to enable accurate mathematical calculations.}
Despite the module undergoing detailed internal deliberation, it finally only outputs the formatted parameters directly applicable to our physics engine.

\section{Experiments}
In this section, we describe the implementation details of our system and demonstrate our evaluation on several datasets. We conduct our experiments aiming to validate that our system can achieve enhanced efficiency while maintaining satisfactory results, and it provides users with an immersive VR environment. All procedures involving human subjects in this research were approved by the Shanghai Jiao Tong University Institutional Review Board for Human Research Protection. All participants provided informed consent prior to participation.

\subsection{Implementations Details}
LIVE-GS focuses on reconstructing an immersive physics-based VR with static Gaussian assets. However, for the completeness of our system, we still integrate the Gaussian training code, which is built on \cite{ye2023gaussian,hu2024semantic}. For our core material analysis module, we construct it with \textit{Langchain} framework \cite{topsakal2023creating}. For physical simulation and interaction in VR, we build our underlying physical system based on Obi solver \cite{obi_solver_website} and \highlightRewrite{Rayfire \cite{rayfire2025}} in Unity 2022.3.30. \highlightRewrite{We also implement a distance field to achieve efficient detection.}

\highlightRewrite{When reconstructing meshes of an elastic body, we choose $\alpha=0.03$. While for rigid body mesh reconstruction, we employed a voxelization parameter of 0.02 combined with $\alpha=0.1$. The sampling density $d$ for particles is set as 15, and the substep is fixed at 4. All other parameters remain constant during the whole experiment.} 

Our VR system is tested on Intel(R) Core(TM) i9-14900K CPU with 64GB memory and an NVIDIA GeForce RTX 4080 SUPER GPU with 16 GB. We conduct our VR experiments on Quest 3 equipment with controllers.

\subsection{Mass Estimation Evaluation} \label{simulation}
Mass is a fundamental parameter shared by both rigid and elastic simulations, with a particularly strong influence on rigid-body behavior. To evaluate mass estimation accuracy, we selected a representative set of 11 objects from the Amazon Berkeley Objects (ABO) dataset, ranging from 0.45 to 25 kilograms, which are numbered and ranked as illustrated in \cref{fig: mass truth}. These objects represent common daily items within typical manual handling ranges. \highlightRewrite{We reconstructed each object for 30000 iterations and arranged them within the same scene. Subsequently, we randomly selected test viewpoints that ensure the target objects remain unobstructed and centered within the field of view. The predicted results from these viewpoints were compared against the ground truth to calculate the corresponding error metrics.} 

Our experimental results are shown in \cref{tab: mass estimations}, demonstrating a robust estimation result with a maximum relative error of \highlightRewrite{$\pm 25.0\%$} and a mean absolute percentage error (MAPE) of \highlightRewrite{$\pm 15.0\%$}.
\begin{table}[htbp]
\centering
\begin{tabular}{>{\centering}p{0.6cm} *{3}{>{\centering\arraybackslash}p{1.2cm}}}
\toprule
ID  & Ground Truth & Predicted Result & Relative Error \\
\midrule
1     & 0.454 & 0.500 & +10.2\%   \\
2      & 0.680 & 0.600 & -11.8\%    \\
3     & 2.876 & 3.500  & +21.7\%   \\
4      & 5.117 & 5.000  & -2.3\%  \\
5      & 6.468 & 7.000  & +8.2\%  \\
6      & 9.072 & 8.000  & -11.8\%  \\
7      & 12.247 & 15.000  & +22.5\%  \\
8      & 17.500 & 15.000  & -14.3\%  \\
9      & 19.999 &  25.000  & +25.0\%  \\
10      & 21.319 & 25.000   & +17.3\%  \\
11      & 24.948 & 20.000   & -19.8\%  \\
\bottomrule
\end{tabular}
 \caption[]{\highlightRewrite{\textbf{Mass estimation results with LIVE-GS (units in kilograms)}.}}
\label{tab: mass estimations}
\end{table}

\subsection{User Study: Evaluation on Elastic Assets}
% User Study: Evaluation on Elastic Assets
As noted earlier, adjusting the parameters for elastic objects presents more complexity. Consequently, we evaluate the system’s authoring capability through experiments on elastic asset creation. We chose 4 datasets from Physdreamer \cite{zhang2024physdreamer} and conducted a user study, where we invited experienced participants to adjust the parameters. We then generate videos with different parameters to compare their effects.

\textbf{Participants}. We recruited 16 participants for this user study, 8 females and 8 males, aged 20 to 30 (\textit{M}=23.75, \textit{SD}= 2.33), who have 2 to 7 years of experience in computer graphics and physical simulation. None of the participants were aware of the research or the experimental purpose.
\begin{figure*}[h]
    \centering
    \includegraphics[width=\linewidth]{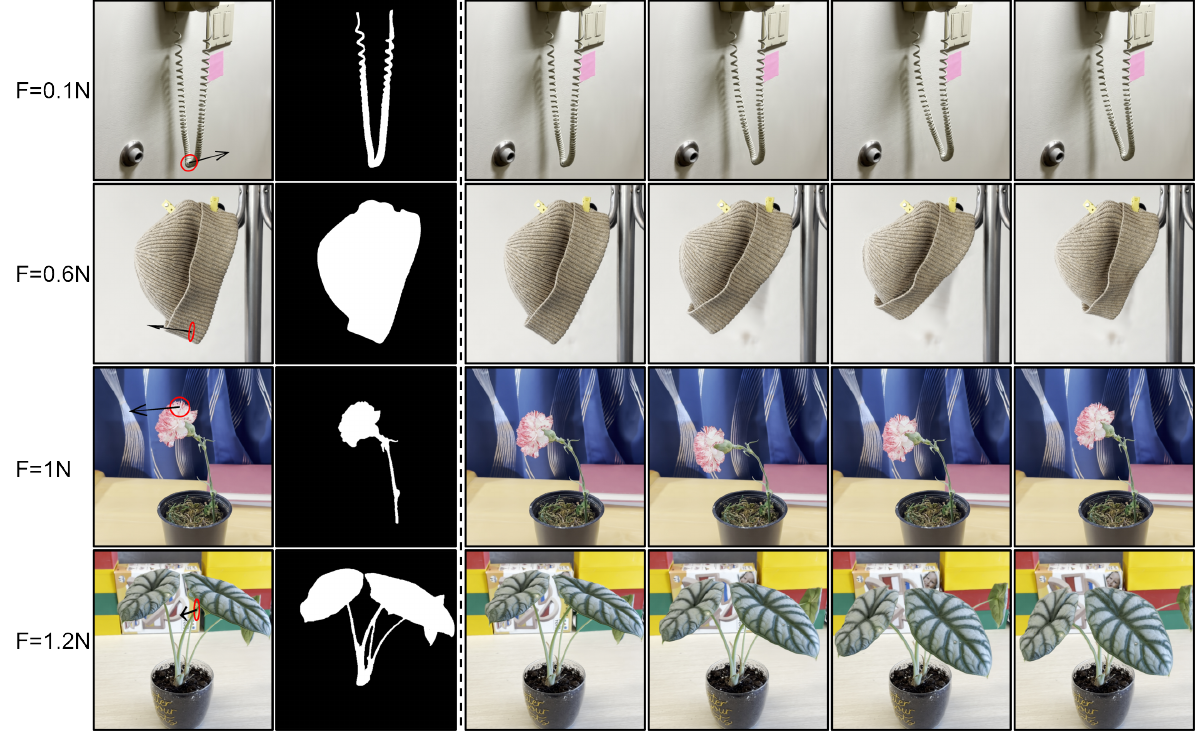}
  \caption{\textbf{Simulation results in \textit{Telephone, Hat, Carnations} and \textit{Alocasia}}. As the description of our system, we use a mask to label our target and analyze the corresponding physical parameters. In each scene, we apply a force for 0.25 seconds, and we show the deformation in 0.5 seconds here.}
\label{fig: Simulation result}
\end{figure*}

\textbf{Setup}. We developed a simple interactive system allowing force application and parameter adjustment via keyboard and mouse. Prior to the experiment, each target underwent preprocessing steps, including segmentation, mesh generation, and particle generation. To ease the difficulty of this task, participants worked with a fixed viewport. During their experiments, they could apply force with different magnitudes and directions to several given positions. The system maintained a rendering frame rate of 30 fps. 

\textbf{Task (about 30 minutes)}. In each trial, participants were presented with a scene with a target elastic object, for which they were required to configure the physical parameters. For more details, they were asked to estimate the \textit{mass} first and then adjust other parameters (deformation resistance and plasticity) to the optimum. We required them to gradually narrow the range until achieving a precision threshold (0.03). Before formal experiments, they first spent 5 to 10 minutes familiarizing themselves with the usage in a study scene. During the experiment, we tracked the users' manual process and recorded relevant data, including time and final parameters. The order of scenes was counter-balanced to prevent order bias.

\textbf{Measurement}. We employed VBench \cite{huang2024vbench,huang2024vbench++}, a comprehensive benchmark suite for assessing video results. We chose 4 dimensions from it, which revealed visual quality (\textit{Image Quality, IQ}), physical realism (\textit{Motion Smoothness, MS}), human perceptions (\textit{Aesthetic Quality, AQ}), and video-text consistency (\textit{Overall Consistency, OC}). The first three dimensions specifically evaluate the quality of generated videos, and the fourth dimension focuses on the semantic alignment of text descriptions and visual content. With users' manual adjustment data \highlightRewrite{and analyzed results (\cref{Tab:prediction result})} of LIVE-GS, we generated videos of the same duration with the consistent settings of viewport, force, and application points. To eliminate errors and randomness, we chose 2 different forces and 2 different points of force application to generate different levels of deformations in each scene, resulting in a total of 4 videos. Each video was accompanied by a standardized text following the format "apply a force of *N to the * for * seconds", corresponding to the actual force application parameters. Additionally, we recorded the time required for manual parameter tuning using the manual approach and our analysis module. For comparative analysis, we included Physdreamer's results utilizing the same Gaussian assets as a baseline benchmark, as it also reconstructs dynamic videos from input images.

\begin{figure}[h]
    \centering
    \includegraphics[width=0.9\linewidth]{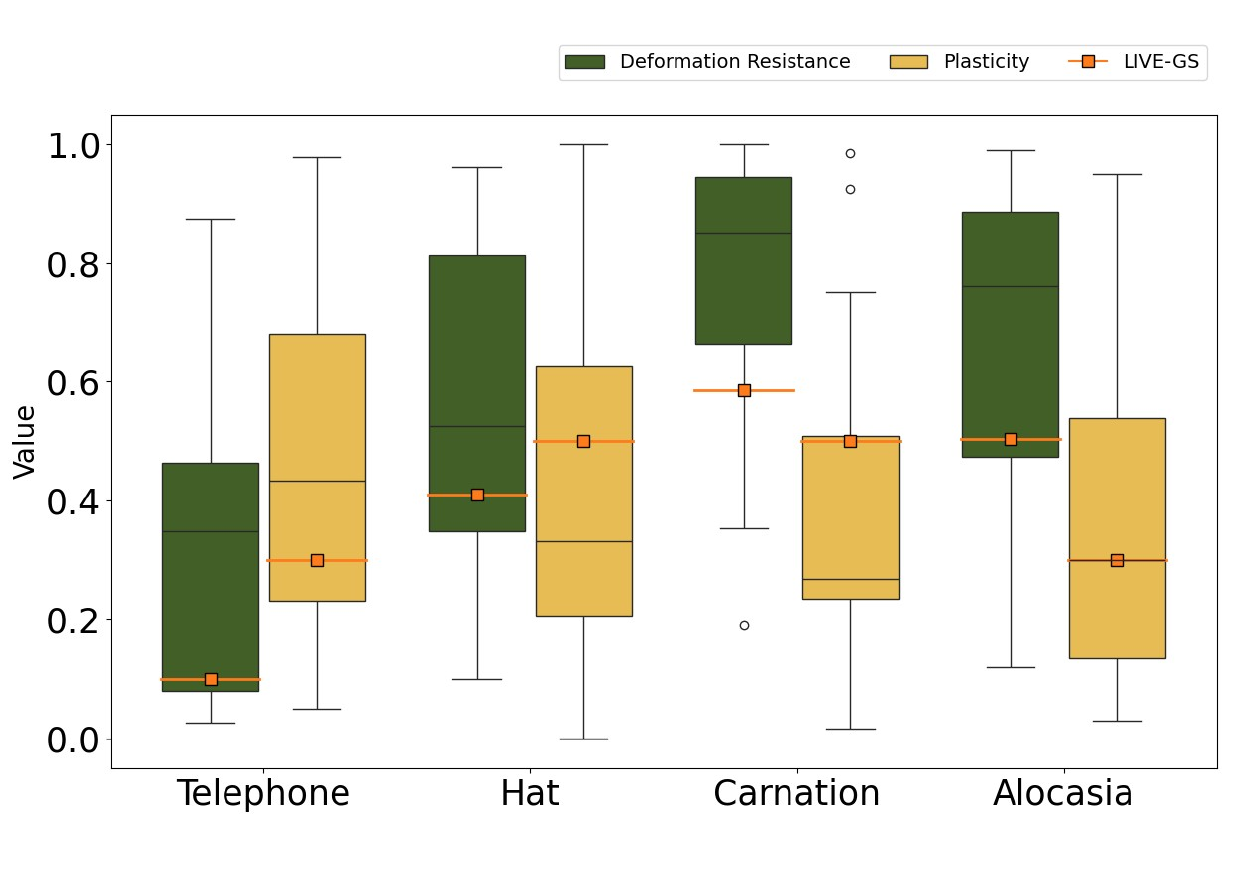}
    \caption{\textbf{Users' manual configurations and LIVE-GS predictions about \textit{Deformation Resistance} and \textit{Plasticity}}. \highlightRewrite{The predicted values of LIVE-GS are also shown here.}}
    \label{fig:userconfig}
\end{figure}

\begin{table}[h]
\centering
\renewcommand{\arraystretch}{1.2}
\begin{tabular}{l >{\centering\arraybackslash}p{0.8cm} >{\centering\arraybackslash}p{1.8cm} >{\centering\arraybackslash}p{1.2cm} }
\toprule
\multicolumn{1}{c}{} & Mass & Deformation Resistance & Plasticity \\
\midrule
% \multicolumn{10}{l}{\textbf{Motion realism}} \\
Telephone     & 0.05 & 0.100  & 0.3   \\
Hat           & 0.15 & 0.411 &  0.5  \\
Carnations     & 0.015 & 0.586  & 0.5   \\
% \multicolumn{10}{l}{\textbf{Visual quality}} \\
Alocasia      & 0.15 & 0.503   & 0.3 \\
\bottomrule
\end{tabular}

\caption{\textbf{Outputs of LIVE-GS for evaluation on Elastic assets (units in kilograms)}.}
\label{Tab:prediction result}
\end{table}

\begin{table*}[!t]
\adjustbox{max width=\textwidth, center}{
\begin{tabular}{@{}cccccccccccccccccccccc@{}}
\toprule[2pt]
\multirow{2}{*}{} & \multirow{2}{*}{\textbf{\begin{tabular}[c]{@{}c@{}}Final\\ Total\end{tabular}}} & \multicolumn{4}{c}{\textbf{Scene Total}} & \multicolumn{4}{c}{\textbf{Motion Smoothness}} & \multicolumn{4}{c}{\textbf{Aesthetic Quality}} & \multicolumn{4}{c}{\textbf{Imaging Quality}} & \multicolumn{4}{c}{\textbf{Overall Consistency}} \\ \cmidrule(l){3-22} 
 &  & Tel. & Hat & Carn. & Aloc. & Tel. & Hat & Carn. & Aloc. & Tel. & Hat & Carn. & Aloc. & Tel. & Hat & Carn. & Aloc. & Tel. & Hat & Carn. & Aloc. \\ \midrule
\multicolumn{1}{c|}{Manual-p1} & \multicolumn{1}{c|}{16} & \highlightA{17} & 15 & 15 & \multicolumn{1}{c|}{8} & 16 & 14 & 11 & \multicolumn{1}{c|}{9} & 16 & 16 & 15 & \multicolumn{1}{c|}{5} & 17 & 15 & 9 & \multicolumn{1}{c|}{15} & 3 & 5 & 7 & 12 \\
\multicolumn{1}{c|}{Manual-p2} & \multicolumn{1}{c|}{5} & 5 & \highlightA{17} & \highlightA{17} & \multicolumn{1}{c|}{1} & 14 & 17 & 10 & \multicolumn{1}{c|}{3} & 2 & 4 & 18 & \multicolumn{1}{c|}{2} & 5 & 18 & 5 & \multicolumn{1}{c|}{1} & 9 & 3 & 15 & 6 \\
\multicolumn{1}{c|}{Manual-p3} & \multicolumn{1}{c|}{18} & 8 & 12 & 11 & \multicolumn{1}{c|}{\highlightA{17}} & 4 & 11 & 8 & \multicolumn{1}{c|}{17} & 8 & 17 & 14 & \multicolumn{1}{c|}{17} & 12 & 10 & 11 & \multicolumn{1}{c|}{8} & 16 & 8 & 2 & 17 \\
\multicolumn{1}{c|}{Manual-p4} & \multicolumn{1}{c|}{11} & 15 & \highlightA{16} & 10 & \multicolumn{1}{c|}{7} & 17 & 9 & 13 & \multicolumn{1}{c|}{10} & 9 & 18 & 8 & \multicolumn{1}{c|}{14} & 4 & 1 & 10 & \multicolumn{1}{c|}{2} & 4 & 18 & 9 & 3 \\
\multicolumn{1}{c|}{Manual-p5} & \multicolumn{1}{c|}{17} & 1 & 2 & \highlightA{18} & \multicolumn{1}{c|}{\highlightA{18}} & 13 & 10 & 15 & \multicolumn{1}{c|}{18} & 3 & 2 & 16 & \multicolumn{1}{c|}{18} & 2 & 9 & 3 & \multicolumn{1}{c|}{17} & 5 & 1 & 17 & 18 \\
\multicolumn{1}{c|}{Manual-p6} & \multicolumn{1}{c|}{15} & 9 & \highlightA{18} & 2 & \multicolumn{1}{c|}{13} & 6 & 18 & 5 & \multicolumn{1}{c|}{12} & 12 & 8 & 2 & \multicolumn{1}{c|}{12} & 9 & 17 & 13 & \multicolumn{1}{c|}{11} & 7 & 4 & 4 & 14 \\
\multicolumn{1}{c|}{Manual-p7} & \multicolumn{1}{c|}{14} & 12 & 14 & \highlightA{16} & \multicolumn{1}{c|}{12} & 7 & 16 & 17 & \multicolumn{1}{c|}{14} & 13 & 7 & 13 & \multicolumn{1}{c|}{6} & 11 & 16 & 4 & \multicolumn{1}{c|}{9} & 14 & 2 & 16 & 13 \\
\multicolumn{1}{c|}{Manual-p8} & \multicolumn{1}{c|}{10} & 14 & 11 & 8 & \multicolumn{1}{c|}{9} & 11 & 13 & 7 & \multicolumn{1}{c|}{8} & 14 & 9 & 12 & \multicolumn{1}{c|}{9} & 10 & 8 & 6 & \multicolumn{1}{c|}{13} & 11 & 6 & 6 & 10 \\
\multicolumn{1}{c|}{Manual-p9} & \multicolumn{1}{c|}{6} & 13 & 4 & 6 & \multicolumn{1}{c|}{10} & 12 & 3 & 6 & \multicolumn{1}{c|}{11} & 15 & 5 & 11 & \multicolumn{1}{c|}{10} & 6 & 5 & 8 & \multicolumn{1}{c|}{12} & 8 & 17 & 3 & 11 \\
\multicolumn{1}{c|}{Manual-p10} & \multicolumn{1}{c|}{12} & \highlightA{18} & 9 & 13 & \multicolumn{1}{c|}{2} & 18 & 8 & 16 & \multicolumn{1}{c|}{5} & 17 & 11 & 17 & \multicolumn{1}{c|}{1} & 18 & 12 & 1 & \multicolumn{1}{c|}{6} & 1 & 11 & 1 & 4 \\
\multicolumn{1}{c|}{Manual-p11} & \multicolumn{1}{c|}{8} & 10 & 13 & 5 & \multicolumn{1}{c|}{6} & 8 & 15 & 1 & \multicolumn{1}{c|}{6} & 10 & 10 & 5 & \multicolumn{1}{c|}{7} & 8 & 13 & 16 & \multicolumn{1}{c|}{14} & 13 & 12 & 14 & 9 \\
\multicolumn{1}{c|}{Manual-p12} & \multicolumn{1}{c|}{13} & 6 & 8 & 12 & \multicolumn{1}{c|}{14} & 1 & 6 & 14 & \multicolumn{1}{c|}{15} & 4 & 13 & 6 & \multicolumn{1}{c|}{11} & 13 & 6 & 7 & \multicolumn{1}{c|}{10} & 18 & 13 & 13 & 16 \\
\multicolumn{1}{c|}{Manual-p13} & \multicolumn{1}{c|}{4} & 7 & 7 & 4 & \multicolumn{1}{c|}{5} & 3 & 7 & 2 & \multicolumn{1}{c|}{2} & 7 & 12 & 4 & \multicolumn{1}{c|}{8} & 14 & 7 & 17 & \multicolumn{1}{c|}{16} & 17 & 10 & 11 & 8 \\
\multicolumn{1}{c|}{Manual-p14} & \multicolumn{1}{c|}{9} & 11 & 6 & 9 & \multicolumn{1}{c|}{11} & 10 & 4 & 12 & \multicolumn{1}{c|}{13} & 11 & 15 & 7 & \multicolumn{1}{c|}{13} & 7 & 4 & 12 & \multicolumn{1}{c|}{5} & 12 & 14 & 12 & 7 \\
\multicolumn{1}{c|}{Manual-p15} & \multicolumn{1}{c|}{1} & 2 & 1 & 7 & \multicolumn{1}{c|}{\highlightA{16}} & 9 & 5 & 4 & \multicolumn{1}{c|}{16} & 6 & 1 & 10 & \multicolumn{1}{c|}{15} & 3 & 2 & 15 & \multicolumn{1}{c|}{7} & 10 & 9 & 8 & 15 \\
\multicolumn{1}{c|}{Manual-p16} & \multicolumn{1}{c|}{2} & 4 & 5 & 3 & \multicolumn{1}{c|}{4} & 2 & 2 & 3 & \multicolumn{1}{c|}{7} & 1 & 14 & 3 & \multicolumn{1}{c|}{4} & 15 & 3 & 14 & \multicolumn{1}{c|}{3} & 15 & 16 & 10 & 2 \\
\multicolumn{1}{c|}{LIVE-GS (ours)} & \multicolumn{1}{c|}{\highlightB{3}} & \highlightB{3} & 10 & 14 & \multicolumn{1}{c|}{\highlightB{3}} & 15 & 12 & 18 & \multicolumn{1}{c|}{\highlightC{4}} & \highlightC{5} & \highlightB{3} & \highlightC{9} & \multicolumn{1}{c|}{\highlightB{3}} & \highlightB{1} & 14 & \highlightB{2} & \multicolumn{1}{c|}{\highlightC{4}} & \highlightC{6} & \highlightC{7} & \highlightC{5} & \highlightC{5} \\
\multicolumn{1}{c|}{PhysDreamer} & \multicolumn{1}{c|}{7} & \highlightA{16} & 3 & 1 & \multicolumn{1}{c|}{15} & 5 & 1 & 9 & \multicolumn{1}{c|}{1} & 18 & 6 & 1 & \multicolumn{1}{c|}{16} & 16 & 11 & 18 & \multicolumn{1}{c|}{18} & 2 & 15 & 18 & 1 \\ \midrule
\multicolumn{1}{c|}{Manual Min Score} & \multicolumn{1}{c|}{73.85\%} & 71.93\% & 71.10\% & 79.24\% & \multicolumn{1}{c|}{71.47\%} & 98.98\% & 97.68\% & 99.61\% & \multicolumn{1}{c|}{97.53\%} & 52.85\% & 49.07\% & 68.45\% & \multicolumn{1}{c|}{59.33\%} & 69.55\% & 74.22\% & 76.60\% & \multicolumn{1}{c|}{67.90\%} & 48.96\% & 33.72\% & 50.36\% & 30.06\% \\
\multicolumn{1}{c|}{Manual Max Score} & \multicolumn{1}{c|}{74.24\%} & 72.83\% & 72.37\% & 79.42\% & \multicolumn{1}{c|}{73.43\%} & 100.00\% & 99.62\% & 99.97\% & \multicolumn{1}{c|}{99.88\%} & 53.71\% & 52.24\% & 68.86\% & \multicolumn{1}{c|}{61.50\%} & 71.13\% & 75.19\% & 76.78\% & \multicolumn{1}{c|}{69.27\%} & 49.60\% & 37.79\% & 51.28\% & 32.67\% \\
\multicolumn{1}{c|}{Dreamer Score} & \multicolumn{1}{c|}{74.09\%} & 72.37\% & 71.68\% & 79.75\% & \multicolumn{1}{c|}{72.55\%} & 99.98\% & 99.72\% & 99.85\% & \multicolumn{1}{c|}{99.94\%} & 52.63\% & 50.04\% & 70.58\% & \multicolumn{1}{c|}{60.56\%} & 70.18\% & 74.44\% & 76.32\% & \multicolumn{1}{c|}{67.09\%} & 49.58\% & 35.02\% & 49.73\% & 32.83\% \\ \bottomrule[2pt]
\end{tabular}
}
\caption{\textbf{Quantitative results of our elastic evaluation experiments}. We list the ranks among LIVE-GS, Physdreamer, and the users' manual results. The \colorbox{yellow}{top 20\% (top 3)} and \colorbox{pink}{top 50\% (top 9)} results are denoted by yellow and pink. Our system ranks 3 out of 18 in \textit{final total score}.}
\label{tab:userstudy1}
\end{table*}

\textbf{Results}. The distribution of participants' parameter configurations for \textit{Deformation Resistance} and \textit{Plasticity} is presented in \cref{fig:userconfig}. For mass configurations, the means and standard deviations for the four scenarios are (units in kilograms): Tel. (\textit{M}=0.0798,\textit{SD}=0.0720); Hat. (\textit{M}=0.0.1753,\textit{SD}=0.1201); Carn. (\textit{M}=0.0672,\textit{SD}=0.0534); Aloc. (\textit{M}=0.0947,\textit{SD}=0.1205). Our quantitative results are normalized and demonstrated in Tab.~\ref{tab:userstudy1}, which includes rankings across different dimensions accompanied by data extremes. Following the evaluation framework in \cite{huang2024vbench,huang2024vbench++}, we calculated scene total scores using a semantic weight of 1 for \textit{OC} and a quality weight of 4 for \textit{MS, AQ} and \textit{IQ}. The final total scores, representing the overall performance, were computed as the average of individual scene totals. In terms of authoring efficiency, compared to the average manual adjustment time of \textbf{331.27 seconds} (about 5.5 minutes), LIVE-GS only requires \textbf{5-10 seconds} for LLM to analyze. A subset of the simulation results from our system is presented in \cref{fig: Simulation result}.

\textbf{Discussion} The analysis of user configuration data in \cref{fig:userconfig} shows a broad distribution, which reflects a relatively wide acceptance range. Furthermore, the rankings of manual results in \cref{tab:userstudy1} exhibit noticeable instability. Taken together, these results suggest that the task is challenging and that users have limited sensitivity to the parameters. This low sensitivity implies that configurations in a reasonably bounded range can provide satisfactory interaction experiences.

As the ranking results presented in \cref{tab:userstudy1}, LIVE-GS maintains competitive behavior, placing 3rd out of 18 in \textit{Final Total} and achieving close to or above the top 50\% across three scenes. Specifically, the range of \textit{Scene Total} is small (less than 0.2\%) in \textit{Carnations}, demonstrating a minor difference between our method and the others despite ranking poorly. As a baseline comparison, Physdreamer gets a lower ranking of 7th in \textit{Final Total}.

Detailed dimensional results reveal that our method achieves a balanced performance across multiple dimensions. Although Physdreamer excels in \textit{motion smoothness}, it suffers from limited \textit{imaging quality} and unstable outcomes in \textit{aesthetic quality (AQ)} and \textit{overall consistency (OC)}. By contrast, LIVE-GS exhibits stable performance across dimensions, achieving leading visual quality in three scenes and consistently ranking within the top 50\% for both \textit{AQ} and \textit{OC}. These high scores in \textit{AQ} and \textit{OC} reveal the advantage for human perception and realistic representation of force application effects, which aligns with our design objective of excellent human-centric VR interactions.

\subsection{User Study: Evaluation on VR Interactions}
% User Study: Evaluation on VR Interactions
Given the significant difference in users' perception of physical properties, it is crucial to validate the performance of our LLM module. To assess the interaction quality and system usability in practical applications, we developed an interactive VR environment including both rigid and elastic objects, and conducted a comprehensive user study with specific interaction tasks and post-experiment questionnaires.
\begin{figure}[ht]
    \centering
    \includegraphics[width=1.0\linewidth]{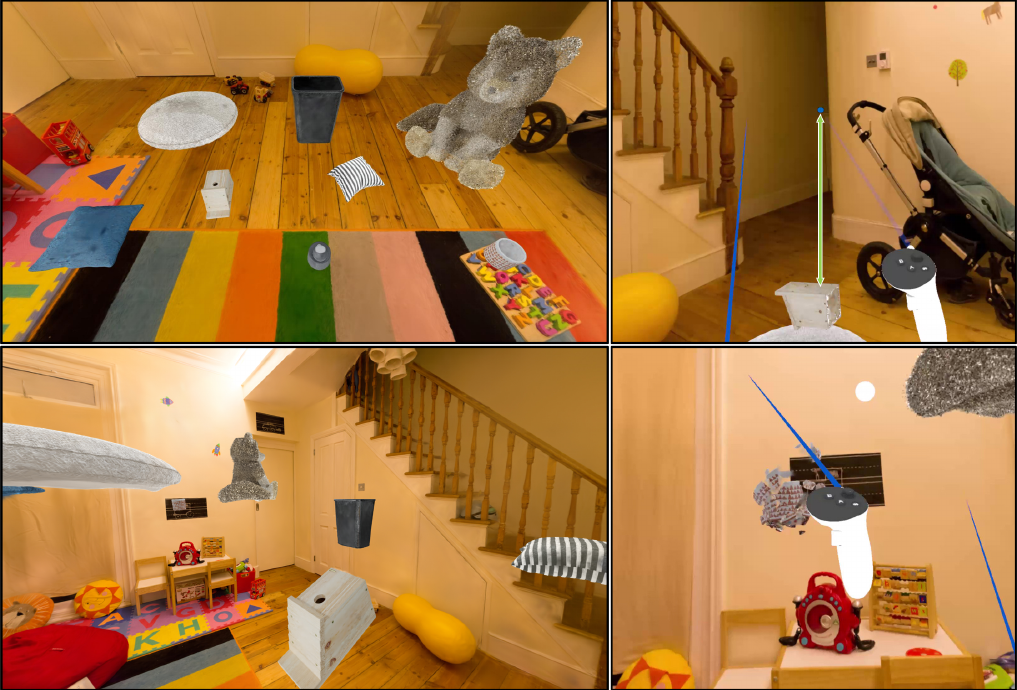}
    \caption{\textbf{Scene for evaluation on VR interactions}. The first row depicts the ray interaction scenario, where users can apply spring forces (indicated by the green arrow). The second row illustrates a shooting interaction scenario, with all objects initialized in mid-air.}
    \label{fig:userstudy ray interaction}
\end{figure}

\textbf{Participants}. We recruited 10 participants for this user study, 5 females and 5 males, aged 19 to 33 (\textit{M}=23.9, \textit{SD}=4.35), consisting of one VR researcher (weekly VR usage), one skilled VR user (monthly VR usage) and 8 non-experts with limited VR experience. None of the participants were aware of the research or the experimental purpose.

\textbf{Setup}. We chose \textit{playroom} from 3DGS \cite{kerbl20233d} as background environment and eight portable objects from ABO and PhysGaussian \cite{xie2024physgaussian}. The objects included 4 elastic and 4 rigid items, carefully selected to represent varying levels of \highlightRewrite{stiffness}, mass, and fragility. The assets were predicted sequentially, with historical information incorporated as context to enhance accuracy. To better visualize the force and physical feedback, we adopt 2 distinct interaction modalities: ray interaction and shoot interaction (\cref{fig:userstudy ray interaction}). The ray interaction modality implemented a spring-based force model at the interaction point, maintaining constant spring stiffness while visually representing force magnitude through proportional displacement. In the other method, participants could shoot balls with identical physical properties, facilitating direct comparison of target objects' physical responses. To ensure low latency, we configured the physics simulation substep at 4, which is sufficient for our computational requirements.

\textbf{Task (about 20 minutes)}. In this experiment, participants were asked to pay attention to the physical feedback when interacting. Observing the high standard deviation in the previous user study suggests that manual parameter estimation is inherently subjective and inconsistent across different users. Consequently, establishing a definitive baseline with specific parameters is impractical. Therefore, this user study forgoes a manually tuned baseline and instead focuses on the plausibility and realism of the parameters generated by our LLM module. Before the main experiment, they underwent a guidance session for about 3 to 5 minutes to familiarize themselves with the usage. In the formal experiment, we designed 2 tasks responding to 2 interaction methods. In the first task, all objects were scattered haphazardly on the ground at initialization, where participants tidied up the messy room, classifying elastic and rigid objects, and organizing them into designated locations. The second task involved shooting fixed midair objects and knocking them down. After each task, participants took a short break and completed specific questionnaires, followed by a comprehensive system usability assessment at the experiment's conclusion. We further interviewed each participant for 5-10 minutes, gathering their feedback and suggestions.

\textbf{Measurement.} Our questionnaire covers subjective tendencies on physical properties and system usability. The task-specific part evaluated participants' perception of physical properties through five ray interaction questions (rigid object mass, elastic object mass, overall mass perception, elastic deformation, and elastic-rigid consistency with expectations) and two shoot interaction questions (physical feedback for elastic and rigid objects). System usability was measured using the standardized System Usability Scale  \cite{bangor2009determining}, comprising ten comprehensive questions.

\begin{figure}[h]
    \centering
    \includegraphics[width=1.0\linewidth]{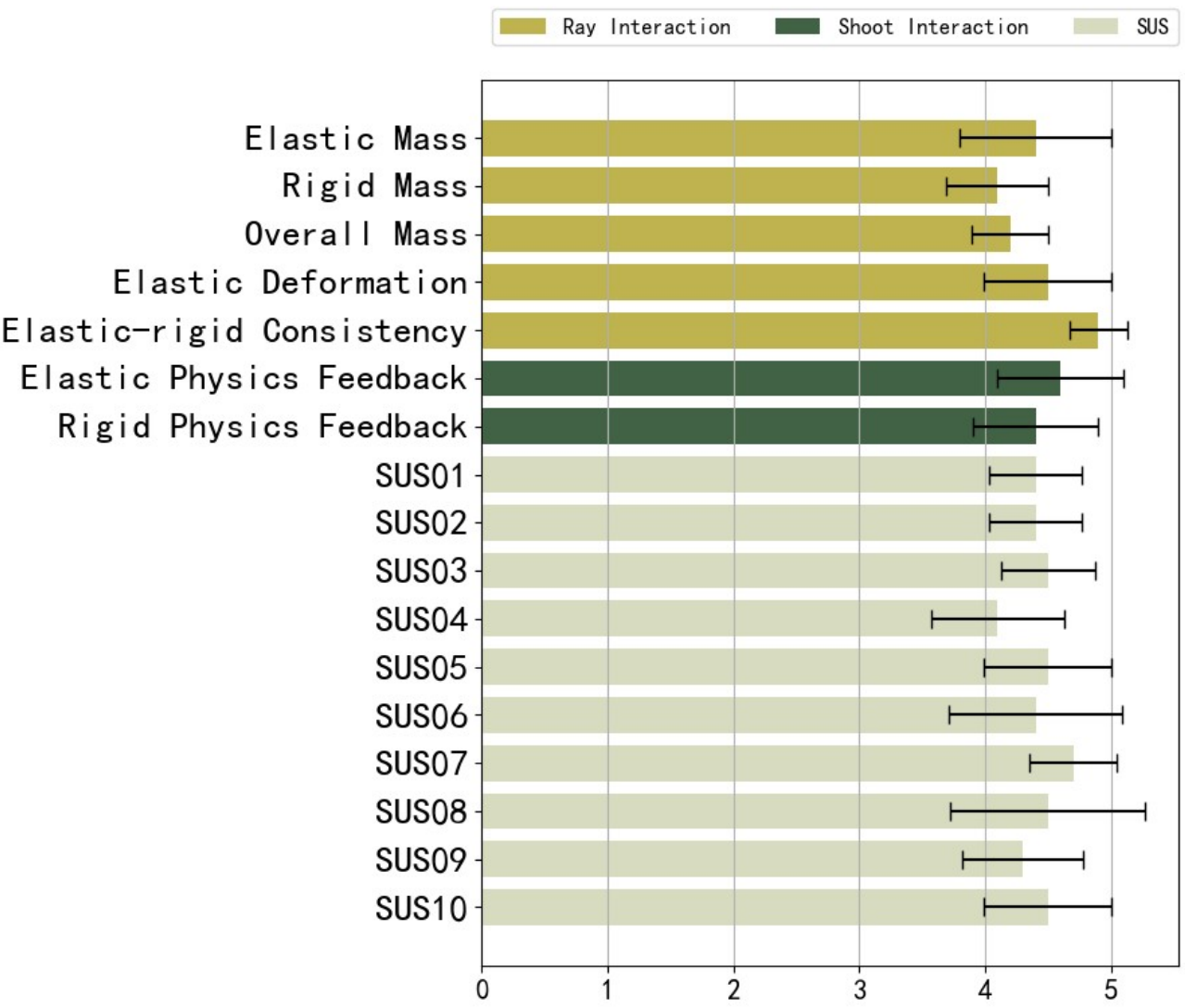}
    \caption{\textbf{Results of the evaluation on VR interactions}. The bar on each histogram represents ±95\% CI. LIVE-GS has received positive opinions from both experienced VR users and beginners.}
    \label{fig:userStudy2_result}
\end{figure}

\textbf{Results}. Our system still requires 5-10 seconds to analyze each object, comparable to the duration in User Study 1. Regarding interaction quality, our system received positive opinions on physical analysis results and system usability. With a System Usability Scale (SUS) score of 88.6, LIVE-GS provides an excellent VR experience according to \cite{bangor2009determining} as illustrated in \cref{fig:userStudy2_result}.

\textbf{Discussion}. Our results indicate a steady analysis time per object, validating the robustness of our framework against scene scale and complexity. Since our analysis module utilizes an image-based analysis with fixed resolution, its processing load is independent of the scene scale. Furthermore, although the historical context accumulates in multi-object tasks, the parallel mechanism at the input stage maintains a nearly constant encoding time, which is negligible compared with the bottleneck caused by token decoding and network transmission. Consequently, the time scales linearly with the number of objects.

Regarding interaction quality, our system got a score above 4 on each question, labeling the excellent interaction quality and functionality. In the post-experiment interview, most participants were satisfied with the physical feedback, especially surprised by the fracture phenomenon. P1 (a non-expert user) commented, "It is amazing that I can destroy the cup! I can't bear to ruin it in reality, and this is so fun!" And P9 (a VR researcher) was impressed by the creativity of LIVE-GS, "This environment shows deeper immersion than traditional modeling methods. I even want to try fluid simulation in my own research with this system!"

\section{Conclusion}
In this paper, we introduce LIVE-GS, a highly realistic interactive GS-based VR system with advanced scene understanding capabilities powered by LLMs. Designed for human-centric interactions in VR, our system effectively meets users’ requirements, delivering physics-aware assets and supporting real-time VR interactions. As Gaussian representations become increasingly prevalent and asset libraries expand, LIVE-GS demonstrates the promising potential of combining 3DGS and LLM for scalable, immersive VR content creation.

\textbf{Limitations and Future Work.} Although LIVE-GS offers a satisfactory VR experience, it remains to be improved. First, the current method relies on a closed-source GPT model without fine-tuning and employs simplified mapping functions to facilitate final outputs. Although current approaches are adequate for VR interactions, they may be incapable of performing precise physical tasks or inferring the properties of uncommon objects. Besides, it may generate unexpected results when analyzing from an inappropriate viewport, which may require a few attempts. While we currently maintain object centrality in the input images and avoid occlusions to ensure reliable predictions, incorporating mechanisms to relax these constraints can improve the system's robustness. For physical simulation, we only implement a simplified presentation of the fracture phenomenon, resulting in blurred edges on fragments. Lastly, our validation regarding elastic parameters is indirect, and more controlled experiments with canonical setups or known parameters can strengthen the physical credibility.

Our future work will focus on developing a domain-specific large model optimized for broader materials and more detailed parameters. We also plan to explore incorporating additional information sources, such as video inputs and multisensory channels (e.g., tactile information), to further enrich the system's capabilities and realism. We will also investigate strategies to improve the robustness of our model to viewpoint changes. More validations, such as including more baselines and different viewpoints, will be added to ensure the robustness of our system.

%% if specified like this the section will be omitted in review mode
\acknowledgments{%
	We thank the anonymous reviewers for their valuable feedback. We thank Hui Wang, Zhi Wang, and Yue Wang for their insightful suggestions. This work was supported by the National Natural Science Foundation of China (62272305) and the National Key Research and Development Program of China (2018YFB1004902).
}

\bibliographystyle{abbrv-doi-hyperref}

\bibliography{template}

\appendix % You can use the `hideappendix` class option to skip everything after \appendix

\end{document}